\documentclass[journal]{IEEEtran}

\IEEEoverridecommandlockouts                              
\title{Relative Coobservability in Decentralized Supervisory Control of Discrete-Event Systems}

\author{Kai Cai, Renyuan Zhang, and W.M. Wonham 
\thanks{K. Cai is with Urban Research Plaza, Osaka City University, Japan.
R. Zhang is with Department of Traffic and Control Engineering,
Northwestern Polytechnical University, China. W.M. Wonham is with
Department of Electrical and Computer Engineering, University of Toronto, Canada. %
Emails: kai.cai@eng.osaka-cu.ac.jp, ryzhang@nwpu.edu.cn,
wonham@control.utoronto.ca. This work was supported in part by
Program to Disseminate Tenure Tracking System, MEXT, Japan; the National Nature Science
Foundation of China, Grant no. 61403308; the Natural Sciences and
Engineering Research Council, Canada, Grant no. 7399.} }

\usepackage{graphicx}
\usepackage{amsthm,amssymb,amscd,amssymb,amsmath,mathrsfs}
\usepackage[sort,compress]{cite}
\usepackage[noend]{algpseudocode}
\usepackage{multirow}
\usepackage{color}
\usepackage{epstopdf}

\newtheorem{thm}{Theorem}

\newtheorem{defn}{Definition}

\newtheorem{prop}{Proposition}

\newtheorem{rem}{Remark}

\begin{document}


\maketitle


\thispagestyle{empty} \pagestyle{plain}

\begin{abstract}
We study the new concept of \emph{relative coobservability} in
decentralized supervisory control of discrete-event systems under
partial observation. This extends our previous work on relative
observability from a centralized setup to a decentralized one. A
fundamental concept in decentralized supervisory control is
coobservability (and its several variations); this property is not,
however, closed under set union, and hence there generally does not
exist the supremal element.  Our proposed relative coobservability,
although stronger than coobservability, is algebraically
well-behaved, and the supremal relatively coobservable sublanguage
of a given language exists.  We present an algorithm to compute this
supremal sublanguage.  Moreover, relative coobservability is weaker
than conormality, which is also closed under set union; unlike
conormality, relative coobservability imposes no constraint on
disabling unobservable controllable events.
\end{abstract}


\section{Introduction} \label{Sec1_Intro}

Recently we introduced the new concept of \emph{relative
observability} in supervisory control of discrete-event systems
(DES) under partial observation (see \cite{CaiZhaWon:14TAC} and its
conference precursor \cite{CaiZhaWon:13CDC}; also the timed case
\cite{CaiZhaWon:14WODES}). Relative observability is stronger than
observability, weaker than normality, and preserved under set union;
hence there exists the supremal relatively observable sublanguage of
a given language, which may be effectively computed. Relative
observability is formulated in a centralized setup where a
monolithic supervisor partially observes and controls the plant as a
whole.

In this paper and its conference precursor \cite{CaiZhaWon:ACC15},
we extend relative observability to a \emph{decentralized} setup
where multiple decentralized supervisors operate jointly, each of
which observes and controls only part of the plant. Decentralized
supervisory control is an effective means of managing computational
complexity when DES are large-scale (e.g. \cite[Chapter~4]{SCDES}).
Our work is motivated by the fact that, in decentralized control
under partial observation, there has so far lacked an effective
concept for which the supremal decentralized supervisors may be
computed, unless normality constraints are imposed which might be
overly conservative.

The fundamental concept in decentralized supervisory control is
\emph{coobservability}, identified in \cite{RudWon:92} (see also
\cite{Cieslak:88}): coobservability and controllability of a
language $K$ is necessary and sufficient for the existence of
\emph{nonblocking} decentralized supervisors that synthesize $K$.
Here the decentralized supervisors follow a \emph{conjunctive}
decision fusion rule: an event is enabled if and only if \emph{all}
supervisors `agree' to enable that event. One may also consider
alternative fusion rules, e.g. that of \emph{disjunctive}, or a mix
of conjunctive and disjunctive; these lead to variations of
coobservability studied in \cite{YooLaf:02}. A further extension
called conditional coobservability is reported in \cite{YooLaf:04a}.

None of the above various versions of coobservability, however, is
closed under set union; consequently there generally does not exist
the supremal coobservable sublanguage of a given language. In fact,
even the existence of a coobservable sublanguage is undecidable in
general \cite{Tri:04letter}. On the other hand, \emph{conormality}
(or strong decomposability), being stronger than coobservability, is
proposed in \cite{RudWon:92}; it is preserved under set union and
the supremal conormal sublanguage may be computed. Conormality,
however, imposes the constraint that no decentralized supervisor can
disable its unobservable, controllable events, and may therefore be
overly conservative in practice. There is a weaker version of
conormality studied in \cite{TakaiUshio:02}, which is also closed
under set union; however, no algorithm is presented to compute the
supremal element.

In this paper, we introduce the new concept of \emph{relative
coobservability}, which is a natural extension of relative
observability to the decentralized supervisory control setup. We
prove that relative coobservability is stronger than (any of the
known variations of) coobservability, weaker than (weak)
conormality, and closed under set union. Moreover, we present an
algorithm for computing the supremal relatively coobservable (and
controllable, $L_m({\bf G})$-closed) sublanguage of a given
language. This algorithm is so far the only one that
effectively synthesizes nonblocking controlled behavior that is generally
more permissive than the conormal counterpart. The new concept
and algorithm are demonstrated with a
Guideway example.



\begin{figure}[!t]
  \centering
  \includegraphics[width=0.4\textwidth]{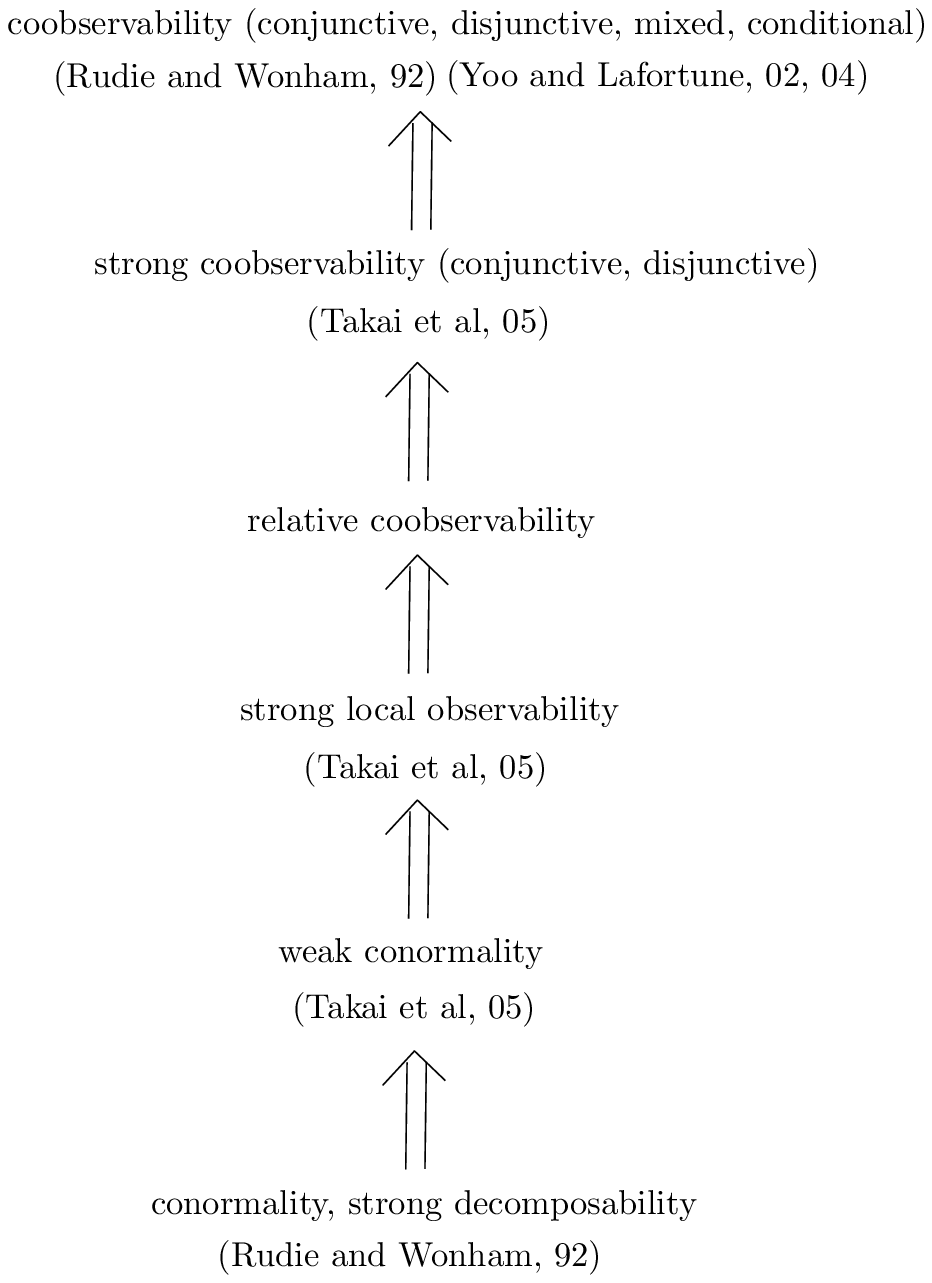}
  \caption{ Observability concepts and their relations
  in decentralized supervisory control under partial observation: bottom to top,
  strong to weak. For all coobservability concepts weaker than relative
  coobservability, no effective algorithm exists that computes
  the corresponding nonblocking controlled behavior.}
  \label{fig:Concepts}
\end{figure}

We note that \cite{TakKumUsh:05} introduced three concepts called
strong conjunctive coobservability, strong disjunctive
coobservability, and strong local observability; the latter two are
proved to be closed under set union. First, for strong local
observability, we will see that it is in fact a special case of our
relative coobservability. Then for strong disjunctive
coobservability, although weaker than our relative coobservability,
there is no existing finitely convergent algorithm that computes its
supremal element. By contrast, we will present an algorithm that
effectively computes the supremal relatively coobservable
sublanguage. The relations of relative coobservability
and other concepts reported in decentralized supervisory control are
summarized in Fig.~\ref{fig:Concepts}.

Note also that, for prefix-closed languages, several procedures are
developed to compute maximal decentralized supervisors, e.g.
\cite{KozWon:95,RohLaf:03}. Those procedures are not, however,
applicable to non-closed languages, because the resulting
decentralized supervisors may be blocking.

Finally we point out that the supremal relatively coobservable
sublanguage of a given language $K$ may be empty even if there
exists a nonempty coobservable sublanguage of $K$ (whether or not
$K$ is prefix-closed). Nevertheless, whenever the supremal
relatively coobservable sublanguage is nonempty (and therefore can
be computed by our proposed algorithm), it is guaranteed to be
coobservable, and nonblocking decentralized supervisors may be
constructed accordingly \cite{RudWon:92}.

%

The rest of the paper is organized as follows. In
Section~\ref{Sec2_RelCoobs} we introduce the new concept of relative
coobservability and show that it is stronger than coobservability
(and its variations) and weaker than conormality. In
Section~\ref{Sec3_SupRelCoobs} we prove that relative
coobservability is closed under set union, and present an algorithm
to compute the supremal relatively coobservable sublanguage of a
given language. The results are demonstrated with a Guideway example
in Section~\ref{Sec4_Guideway}.
Finally in Section~\ref{Sec5_Concl} we state our conclusions.

\section{Relative Coobservability} \label{Sec2_RelCoobs}

The plant to be controlled is modeled by a generator
\begin{align} \label{eq:generator}
\textbf{G} = (Q, \Sigma, \delta, q_0, Q_m)
\end{align}
where $Q$ is the finite state set; $q_0 \in Q$ the initial state;
$Q_m \subseteq Q$ the subset of marker states; $\Sigma$ the finite
event set; $\delta: Q \times \Sigma \rightarrow Q$ the (partial)
state transition function. In the usual way, $\delta$ is extended to
$\delta: Q \times \Sigma^* \rightarrow Q$, and we write
$\delta(q,s)!$ to mean that $\delta(q,s)$ is defined. The
\emph{closed behavior} of $\textbf{G}$ is the language
\begin{align} \label{eq:closedlang}
L(\textbf{G}) := \{s \in \Sigma^* | \delta(q_0,s)!\} \subseteq
\Sigma^*
\end{align}
and the \emph{marked behavior} is
\begin{align} \label{eq:markedlang}
L_m(\textbf{G}) := \{s \in L(\textbf{G}) | \delta(q_0,s) \in Q_m\}
\subseteq L(\textbf{G}).
\end{align}
A string $s_1$ is a \emph{prefix} of a string $s$, written $s_1 \leq
s$, if there exists $s_2$ such that $s_1 s_2 = s$. The
\emph{(prefix) closure} of $L_m(\textbf{G})$ is
$\overline{L_m(\textbf{G})} := \{ s_1 \in \Sigma^* \ |\ (\exists s
\in L_m(\textbf{G})) s_1 \leq s \}$. In this paper we assume
$\overline{L_m(\textbf{G})} = L(\textbf{G})$; namely $\textbf{G}$ is
\emph{nonblocking}. A language $K \subseteq \Sigma^*$ is
$L_m(\textbf{G})$\emph{-closed} if $\overline{K} \cap
L_m(\textbf{G}) = K$.

For partial observation, let the event set $\Sigma$ be partitioned
into $\Sigma_o$, the observable event subset, and $\Sigma_{uo}$, the
unobservable subset (i.e. $\Sigma = \Sigma_o \dot{\cup}
\Sigma_{uo}$). Bring in the \emph{natural projection} $P : \Sigma^*
\rightarrow \Sigma_o^*$ defined according to
\begin{equation} \label{eq:natpro}
\begin{split}
P(\epsilon) &= \epsilon, \ \ \epsilon \mbox{ is the empty string;} \\
P(\sigma) &= \left\{
  \begin{array}{ll}
    \epsilon, & \hbox{if $\sigma \notin \Sigma_o$,} \\
    \sigma, & \hbox{if $\sigma \in \Sigma_o$;}
  \end{array}
\right.\\
P(s\sigma) &= P(s)P(\sigma),\ \ s \in \Sigma^*, \sigma \in \Sigma.
\end{split}
\end{equation}
In the usual way, $P$ is extended to $P : Pwr(\Sigma^*) \rightarrow
Pwr(\Sigma^*_o)$, where $Pwr(\cdot)$ denotes \emph{powerset}. Write
$P^{-1} : Pwr(\Sigma^*_o) \rightarrow Pwr(\Sigma^*)$ for the
\emph{inverse-image function} of $P$.


Let $\Sigma_{o,i} \subseteq \Sigma$ and the natural projections $P_i
: \Sigma^* \rightarrow \Sigma_{o,i}^*$, $i \in \mathcal {I}$
($\mathcal {I}$ is some index set).  Also let $\Sigma_{c,i}
\subseteq \Sigma$. We consider decentralized supervisory control
where each decentralized supervisor $i \in \mathcal {I}$ observes
events only in $\Sigma_{o,i}$, and controls events only in
$\Sigma_{c,i}$.  Then let $\Sigma_c := \cup_{i \in \mathcal {I}}
\Sigma_{c,i}$ be the total controllable event subset, and $\Sigma_u
:= \Sigma \setminus \Sigma_c$ the uncontrollable subset. A language
$K \subseteq \Sigma^*$ is \emph{controllable} with respect to {\bf
G} if
\begin{align}
\overline{K}\Sigma_u \cap L({\bf G}) \subseteq \overline{K}.
\end{align}

For conceptual simplicity let us first consider the case of two
decentralized supervisors, i.e. $\mathcal {I} = \{1,2\}$. The
(conjunctive) coobservability is defined as follows
\cite{RudWon:92}. A language $K \subseteq L_m({\bf G})$ is
\emph{coobservable} with respect to ${\bf G}$, $P_1$, $P_2$,
$\Sigma_{c,1}$, $\Sigma_{c,2}$ if
\begin{align}
&(\forall s, s', s'' \in \Sigma^*)\ P_1(s)=P_1(s') \wedge
P_2(s)=P_2(s'') \Rightarrow \notag\\
& \mbox{(i) } (\forall \sigma \in \Sigma_{c,1} \cap \Sigma_{c,2}) \notag\\
&\hspace{1cm} (s'\sigma \in \overline{K} \wedge s \in \overline{K}
\wedge s\sigma \in
L({\bf G}) \Rightarrow s\sigma \in \overline{K}) \notag\\
&\hspace{0.6cm} \vee (s''\sigma \in \overline{K} \wedge s \in
\overline{K} \wedge s\sigma \in L({\bf G}) \Rightarrow s\sigma \in \overline{K}) \label{eq:CPcoobs1}\\
& \mbox{(ii) } (\forall \sigma \in \Sigma_{c,1} \setminus
\Sigma_{c,2}) \notag\\
&\hspace{1cm} s'\sigma \in \overline{K} \wedge s \in \overline{K}
\wedge s\sigma \in
L({\bf G}) \Rightarrow s\sigma \in \overline{K} \label{eq:CPcoobs2}\\
& \mbox{(iii) } (\forall \sigma \in \Sigma_{c,2} \setminus
\Sigma_{c,1}) \notag\\
&\hspace{1cm} s''\sigma \in \overline{K} \wedge s \in \overline{K}
\wedge s\sigma \in L({\bf G}) \Rightarrow s\sigma \in \overline{K}
\label{eq:CPcoobs3}
\end{align}

First observe that (ii) (resp. (iii)) above, for a controllable
event $\sigma$ belonging only to $\Sigma_{c,1}$, i.e. $\sigma \in
\Sigma_{c,1} \setminus \Sigma_{c,2}$ (resp. $\sigma \in \Sigma_{c,2}
\setminus \Sigma_{c,1}$), is simply the standard observability
condition \cite{LinWon:88Obs} with respect to $P_1$ (resp. $P_2$)
that is applied. For a shared controllable event $\sigma \in
\Sigma_{c,1} \cap \Sigma_{c,2}$ in (i) above, on the other hand,
both observations $P_1$ and $P_2$ are involved, and the condition
(\ref{eq:CPcoobs1}) is equivalent to
\begin{align*}
s'\sigma \in \overline{K} \wedge s''\sigma \in \overline{K} \wedge s
\in \overline{K} \wedge s\sigma \in L({\bf G}) \Rightarrow s\sigma
\in \overline{K}
\end{align*}
namely the decision of enabling $\sigma$ after string $s$ will be
made if it is first ratified by both supervisors working through
their respective observation channels.

Coobservability, together with controllability and
$L_m(\textbf{G})$-closedness, of a language $K$ is shown to be
necessary and sufficient for the existence of two decentralized
supervisors \emph{conjunctively} synthesizing $K$ \cite{RudWon:92}.
Coobservability, however, is not closed under set union, and
consequently the supremal coobservable sublanguage of $K$ need not
exist in general. This fact motivates us to propose the new concept,
\emph{relative coobservability}, which (as we will show) is
algebraically better behaved.


\begin{defn} \label{defn:relcoobs}
Let $C \subseteq L_m(\textbf{G})$ be a fixed \emph{ambient}
sublanguage. A sublanguage $K \subseteq C$ is \emph{relatively
coobservable}, or simply $\overline{C}$\emph{-coobservable}, with
respect to ${\bf G}$, $P_1$, $P_2$, $\Sigma_{c,1}$, $\Sigma_{c,2}$
if
\begin{align}
&(\forall s, s', s'' \in \Sigma^*)\ P_1(s)=P_1(s') \wedge
P_2(s)=P_2(s'') \Rightarrow \notag\\
& \mbox{(i) } (\forall \sigma \in \Sigma_{c,1} \cap \Sigma_{c,2}) \notag\\
&\hspace{1cm} (s'\sigma \in \overline{K} \wedge s \in \overline{C}
\wedge s\sigma \in
L({\bf G}) \Rightarrow s\sigma \in \overline{K}) \notag\\
&\hspace{0.6cm} \wedge (s''\sigma \in \overline{K} \wedge s \in
\overline{C} \wedge s\sigma \in L({\bf G}) \Rightarrow s\sigma \in \overline{K}) \label{eq:relcoobs1}\\
& \mbox{(ii) } (\forall \sigma \in \Sigma_{c,1} \setminus
\Sigma_{c,2}) \notag\\
&\hspace{1cm} s'\sigma \in \overline{K} \wedge s \in \overline{C}
\wedge s\sigma \in
L({\bf G}) \Rightarrow s\sigma \in \overline{K} \label{eq:relcoobs2}\\
& \mbox{(iii) } (\forall \sigma \in \Sigma_{c,2} \setminus
\Sigma_{c,1}) \notag\\
&\hspace{1cm} s''\sigma \in \overline{K} \wedge s \in \overline{C}
\wedge s\sigma \in L({\bf G}) \Rightarrow s\sigma \in \overline{K}
\label{eq:relcoobs3}
\end{align}
\end{defn}

\begin{figure}[!t]
  \centering
  \includegraphics[width=0.4\textwidth]{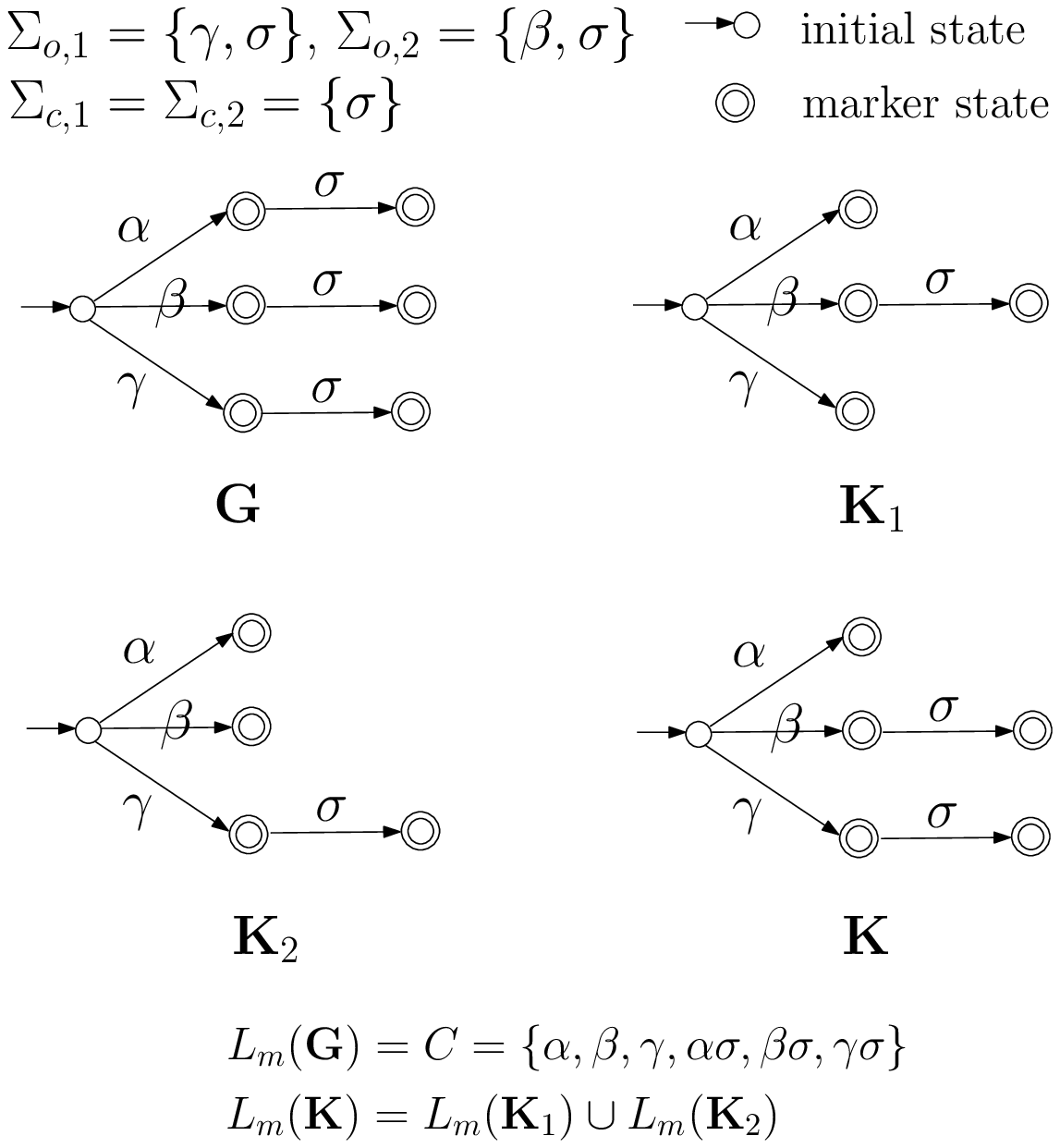}
  \caption{Suppose that $\vee$ were used in (\ref{eq:relcoobs1}). Then $L_m({\bf K}_1)$ and $L_m({\bf K}_2)$ would both be
  $\overline{C}$-coobservable, but the union $L_m({\bf K}) = L_m({\bf K}_1) \cup L_m({\bf
  K}_2)$ would not be. The reason is as follows. First for $P_1: \Sigma^* \rightarrow \Sigma_{o,1}^*$, let $s=\alpha$ and $s'=\beta$. Then $P_1(s)=P_1(s')=\epsilon$,
  $s' \sigma \in L({\bf K})$, $s \in \overline{C}$, $s\sigma \in L({\bf G})$, but $s\sigma \notin L({\bf
  K})$. Second for $P_2: \Sigma^* \rightarrow \Sigma_{o,2}^*$, let $s=\alpha$ and $s''=\gamma$. Then $P_2(s)=P_2(s'')=\epsilon$,
  $s'' \sigma \in L({\bf K})$, $s \in \overline{C}$, $s\sigma \in L({\bf G})$, but $s\sigma \notin L({\bf
  K})$. (Notation: we will use the same initial and marker state notation in subsequent figures.)}
  \label{fig:relcoobs_CPcoobs}
\end{figure}

Several remarks on the definition are in order.  First, relative
coobservability is a `strengthened' version of coobservability in
two respects. For one, all strings $s$ in the ambient $\overline{C}$
are considered, instead of just strings in $\overline{K}$.
For the other, the two implications in (\ref{eq:relcoobs1}) are
connected by ``and'' $\wedge$, instead of ``or'' $\vee$. Namely
(\ref{eq:relcoobs1}) requires that the `observational consistency'
hold for {\it both} observation channels $P_1$ and $P_2$. This
requirement is crucial to provide closure under union for relative
coobservability; as the example in Fig.~\ref{fig:relcoobs_CPcoobs}
shows, using $\vee$ in (\ref{eq:relcoobs1}) would fail to guarantee
closure under union.\footnote{This requirement is admittedly a shortcoming
of our relative coobservability approach as it rules out any inconsistency in
decentralized supervisors' local decisions. However, in the absence of such a
requirement it does not seem possible to preserve the property of closure
under union, and hence the effective computability of a useful result.
Computation of a merely ``maximal'', as distinct from supremal, behavior
(even if that could be achieved) would be, in our view, of little practical interest.}
Hence we have identified the two defects that cause coobservability
to fail to be closed under union: (1) lack of an ambient language,
(2) the use of disjunctive (``or'') $\vee$ logic in connecting local
observational consistency.

The above two (strengthening) modifications lead immediately to the
following.

\begin{prop} \label{prop:relcoobs_CPcoobs}
If $K \subseteq  C$ is $\overline{C}$-coobservable, then $K$ is also
coobservable.
\end{prop}

The reverse statement need not be true.  For an example see again
Fig.~\ref{fig:relcoobs_CPcoobs}: $L_m({\bf K}_1)$ (or $L_m({\bf
K}_2)$) is coobservable (since $\vee$ is used in
(\ref{eq:CPcoobs1})) but not relatively coobservable ($\wedge$ used
in (\ref{eq:relcoobs1})).

Second, relative coobservability is a decentralized version of
relative observability \cite{CaiZhaWon:14TAC}. Indeed, for an
unshared controllable event, namely (ii) and (iii) in the
definition, individual relative observability conditions
corresponding to the respective natural projections are applied;
while for a shared controllable event, namely (i), both conditions
must be satisfied simultaneously. This implies that the definition
of relative coobservability is equivalent to the condition that for
each $i \in \mathcal {I}$, $K$ is relatively observable with respect
to $P_i$, i.e.
\begin{align} \label{eq:relobs}
(\forall s,s' \in \Sigma^*)&(\forall \sigma \in \Sigma_{c,i})\
P_i(s)=P_i(s') \wedge s'\sigma \in
\overline{K} \notag \\
&\wedge s \in \overline{C} \wedge s\sigma \in L({\bf G}) \Rightarrow
s\sigma \in \overline{K}.
\end{align}
Thus we see that Definition~\ref{defn:relcoobs} is easily adapted to
a general finite set $\mathcal {I}$ of decentralized supervisors.
For this reason, we also refer to relative coobservability as
$\mathcal {I}$\emph{-fold relative observability}.

Third, consider a finite set $\mathcal {I}$ of
decentralized supervisors. Relative coobservability ensures that if
a decentralized supervisor enables (resp. disables) an event, then
no other supervisor disables (resp. enables) that event. Namely,
there is no conflict among decentralized supervisors' local control
decisions, and each supervisor may independently decide to enable or
disable an event based on its local observation.

Fourth, we note that the ambient language $C$ is selected such that
all the strings in $\overline{C}$ must be tested for the conditions
of relative coobservability.  In addition, if $C_1 \subseteq C_2
\subseteq L_m(\textbf{G})$ are two ambient languages, it follows
easily from Definition~\ref{defn:relcoobs} that
$\overline{C_2}$-coobservability implies
$\overline{C_1}$-coobservability. Namely, the smaller the ambient
language, the weaker the relative coobservability.

\medskip

An alternative definition of coobservability that has appeared in
the literature is disjunctive coobservability \cite{YooLaf:02},
defined as follows. A language $K \subseteq L_m({\bf G})$ is
\emph{disjunctively coobservable} with respect to ${\bf G}$, $P_1$,
$P_2$, $\Sigma_{c,1}$, $\Sigma_{c,2}$ if
\begin{align}
&(\forall s, s', s'' \in \Sigma^*)\ P_1(s)=P_1(s') \wedge
P_2(s)=P_2(s'') \Rightarrow \notag\\
& \mbox{(i) } (\forall \sigma \in \Sigma_{c,1} \cap \Sigma_{c,2})\ s'\sigma \in L({\bf G}) \setminus \overline{K} \wedge s''\sigma \in L({\bf G}) \setminus \overline{K} \notag\\
&\hspace{1.1cm}  \wedge s \in \overline{K} \wedge s\sigma \in
L({\bf G}) \Rightarrow s\sigma \in L({\bf G}) \setminus \overline{K} \label{eq:DAcoobs1}\\
& \mbox{(ii) } (\forall \sigma \in \Sigma_{c,1} \setminus
\Sigma_{c,2})\ s'\sigma \in L({\bf G}) \setminus \overline{K} \notag\\
&\hspace{1.1cm} \wedge s \in \overline{K} \wedge s\sigma \in
L({\bf G}) \Rightarrow s\sigma \in L({\bf G}) \setminus \overline{K} \label{eq:DAcoobs2}\\
& \mbox{(iii) } (\forall \sigma \in \Sigma_{c,2} \setminus
\Sigma_{c,1})\ s''\sigma \in L({\bf G}) \setminus \overline{K} \notag\\
&\hspace{1.1cm} \wedge s \in \overline{K} \wedge s\sigma \in L({\bf
G}) \Rightarrow s\sigma \in L({\bf G}) \setminus \overline{K}
\label{eq:DAcoobs3}
\end{align}
Disjunctive coobservability requires that for a shared controllable
event $\sigma$ in (i) above, the decision of disabling $\sigma$
after string $s$ be ratified by both supervisors working through
their respective observation channels. This implies that $\sigma$
will be enabled if some supervisor decides to enable it, therefore
the name ``disjunctive''. Disjunctive coobservability is different
from conjunctive coobservability, and in general neither of the two
versions implies the other \cite{YooLaf:02}.

Disjunctive coobservability, together with controllability and
$L_m(\textbf{G})$-closedness, of a language $K$ is proved to be
necessary and sufficient for the existence of two decentralized
supervisors \emph{disjunctively} synthesizing $K$ \cite{YooLaf:02}.
Again, however, it is not closed under set union, and consequently
the supremal element need not exist in general. We show next that
our relative coobservability is stronger than disjunctive
coobservability.

\begin{prop} \label{prop:relcoobs_DAcoobs}
If $K \subseteq  C$ is $\overline{C}$-coobservable, then $K$ is also
disjunctively coobservable.
\end{prop}

\emph{Proof.} Let $s, s', s'' \in \overline{K} \subseteq
\overline{C}$, $P_1(s)=P_1(s')$, and $P_2(s)=P_2(s'')$. We show that
condition (i), namely (\ref{eq:DAcoobs1}), of disjunctive
coobservability holds. Let $\sigma \in \Sigma_{c,1} \cap
\Sigma_{c,2}$, $s'\sigma \in L({\bf G}) \setminus \overline{K}$,
$s''\sigma \in L({\bf G}) \setminus \overline{K}$, and $s\sigma \in
L({\bf G})$. We will show that $s\sigma \in L({\bf G}) \setminus
\overline{K}$. From (\ref{eq:relcoobs1}) we know that
\begin{align*}
&(s'\sigma \notin \overline{K} \Rightarrow s'\sigma \notin L({\bf
G}) \vee s' \notin \overline{C} \vee s\sigma \notin \overline{K})\\
\wedge \ &(s''\sigma \notin \overline{K} \Rightarrow s''\sigma
\notin L({\bf G}) \vee s'' \notin \overline{C} \vee s\sigma \notin
\overline{K}).
\end{align*}
We have $s'\sigma \notin \overline{K}$, $s'\sigma \in L({\bf G})$,
$s' \in \overline{C}$; and $s''\sigma \notin \overline{K}$,
$s''\sigma \in L({\bf G})$, $s'' \in \overline{C}$. It follows that
$s\sigma \notin \overline{K}$. Since $s\sigma \in L({\bf G})$, we
conclude that $s\sigma \in L({\bf G}) \setminus \overline{K}$.

The same reasoning proves conditions (ii) and (iii), namely
(\ref{eq:DAcoobs2}) and (\ref{eq:DAcoobs3}), of disjunctive
coobservability.\footnote{That relative coobservability (or
$\mathcal {I}$-fold relative observability) is stronger than
disjunctive coobservability
(Proposition~\ref{prop:relcoobs_DAcoobs}) or conjunctive
coobservability (Proposition~\ref{prop:relcoobs_CPcoobs}) can also
be proved by noting that it is stronger than a property called local
observability \cite{TakKumUsh:05}: local observability requires that
for each $i \in \mathcal {I}$, $K$ be observable with respect to
$P_i$, i.e. $\mathcal {I}$-fold observability, and is proved to be
stronger than disjunctive and conjunctive coobservability.} \hfill
$\square$

\begin{figure}[!t]
  \centering
  \includegraphics[width=0.5\textwidth]{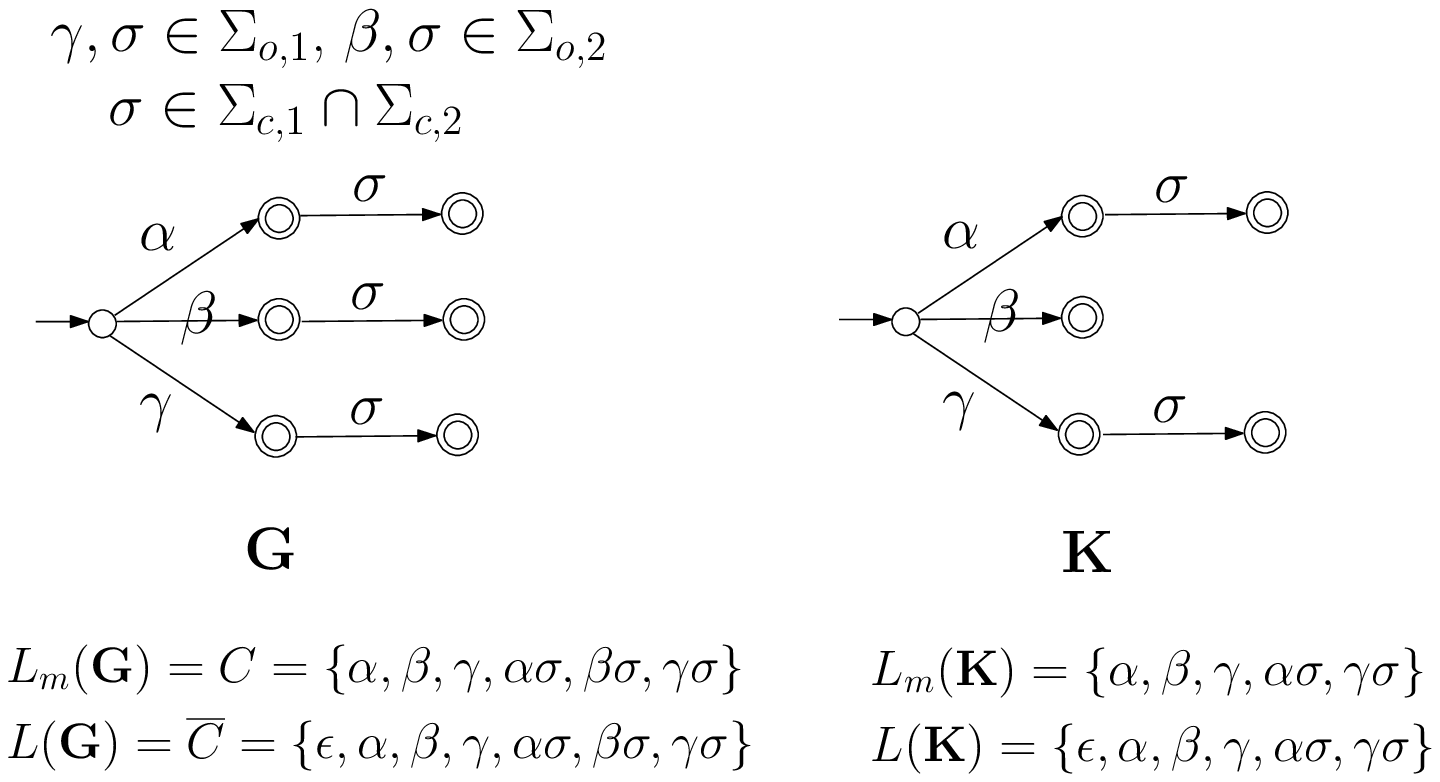}
  \caption{$L_m(\textbf{K})$ is disjunctively coobservable but not relatively coobservable.
  For $P_1$, let $s=\beta$ and $s'=\alpha$. Then $P_1(s)=P_1(s')=\epsilon$,
  $s' \sigma \in L({\bf K})$, $s \in \overline{C}$, $s\sigma \in L({\bf G})$,
  but $s\sigma \notin L({\bf K})$. This violates (\ref{eq:relcoobs1}),
  and therefore relative coobservability fails. For $P_2$, on the other hand, let $s''=\gamma$ so that $P_2(s')=P_2(s'')=\epsilon$.
  The fact that $s''\sigma \in L({\bf K})$ makes (\ref{eq:DAcoobs1}) true.
  One may check that disjunctive coobservability of $L_m(\textbf{K})$ indeed holds.}
  \label{fig:relcoobs_DAcoobs}
\end{figure}

The reverse statement of Proposition~\ref{prop:relcoobs_DAcoobs}
need not be true.  An example is displayed in
Fig.~\ref{fig:relcoobs_DAcoobs}, of a disjunctively coobservable
language that is not relatively coobservable.

\begin{rem}
We note that in \cite{TakKumUsh:05}, ``strong conjunctive'' and
``strong disjunctive'' coobservability are studied, the essence
being to choose strings from the ambient language $L_m({\bf G})$
instead of $K$. For that reason they are stronger than their
respective type of coobservability. Strong disjunctive
coobservability is shown to be closed under set union (while strong
conjunctive coobservability is not), but no finitely convergent
algorithm is given to compute the supremal element. Our relative
coobservability may be shown to be stronger than these strong
versions of coobservability; nevertheless we shall present an
algorithm that computes the supremal relatively coobservable
sublanguage of a given language.

We also note in passing that since either conjunctive or disjunctive
coobservability is stronger than the mixed coobservability
\cite{YooLaf:02}, which is furthermore stronger than the conditional
coobservability \cite{YooLaf:04a}, our coobservability is stronger
than all versions of coobservability reported in the literature.
\end{rem}

\medskip

We turn now to prove that relative coobservability is weaker than
conormality (or strong decomposibility in \cite{RudWon:92}). A
language $K \subseteq L_m({\bf G})$ is \emph{conormal} with respect
to ${\bf G}$, $P_1$, $P_2$, $\Sigma_{c,1}$, $\Sigma_{c,2}$ if
\begin{align} \label{eq:conorm}
\left( P^{-1}_1 P_1 (\overline{K}) \cup P^{-1}_2 P_2 (\overline{K})
\right) \cap L({\bf G}) = \overline{K}.
\end{align}
Conormality may be overly restrictive because it requires that for
each decentralized supervisor $i \in \mathcal {I}$, only observable
(under $P_i$), controllable events may be disabled. Relative
coobservability, by contrast, does not impose this restriction, i.e.
control may be exercised by each decentralized supervisor over its
unobservable controllable events.

\begin{prop} \label{prop:relcoobs_norm}
If $K \subseteq C$ is conormal with respect to ${\bf G}$, $P_1$,
$P_2$, $\Sigma_{c,1}$, $\Sigma_{c,2}$, then $K$ is
$\overline{C}$-coobservable.
\end{prop}
\emph{Proof.} Let $s,s',s'' \in \Sigma^*$, $P_1(s) = P_1(s')$, and
$P_2(s) = P_2(s'')$. We show that
(\ref{eq:relcoobs1})-(\ref{eq:relcoobs3}) all hold. First for
(\ref{eq:relcoobs1}), let $\sigma \in \Sigma_{c,1} \cap
\Sigma_{c,2}$, $s' \sigma \in \overline{K}$, $s \in \overline{C}$,
and $s \sigma \in L(\textbf{G})$; it will be shown that $s \sigma
\in \overline{K}$. From $s' \sigma \in \overline{K}$ we have
\begin{align*}
P_1(s' \sigma) \in P_1\overline{K} &\Rightarrow P_1(s) P_1(\sigma)
\in P_1\overline{K} \\
&\Rightarrow s\sigma \in P_1^{-1}P_1\overline{K} \\
&\Rightarrow s\sigma \in P^{-1}_1 P_1 (\overline{K}) \cup P^{-1}_2
P_2 (\overline{K})
\end{align*}
Hence $s\sigma \in \left( P^{-1}_1 P_1 (\overline{K}) \cup P^{-1}_2
P_2 (\overline{K}) \right) \cap L({\bf G}) = \overline{K}$ by
conormality of $\overline{K}$. Similarly, let $s'' \sigma \in
\overline{K}$; through $P_2$ we derive $s\sigma \in \overline{K}$.

For (\ref{eq:relcoobs2}), let $\sigma \in \Sigma_{c,1} \setminus
\Sigma_{c,2}$, $s' \sigma \in \overline{K}$, $s \in \overline{C}$,
and $s \sigma \in L(\textbf{G})$. By the same derivation as above,
we get $s\sigma \in \overline{K}$. Finally for (\ref{eq:relcoobs3}),
let $\sigma \in \Sigma_{c,2} \setminus \Sigma_{c,1}$, $s'' \sigma
\in \overline{K}$, $s \in \overline{C}$, and $s \sigma \in
L(\textbf{G})$. Again by the same derivation as above but through
$P_2$, we get $s\sigma \in \overline{K}$. \hfill $\square$

\begin{figure}[!t]
  \centering
  \includegraphics[width=0.5\textwidth]{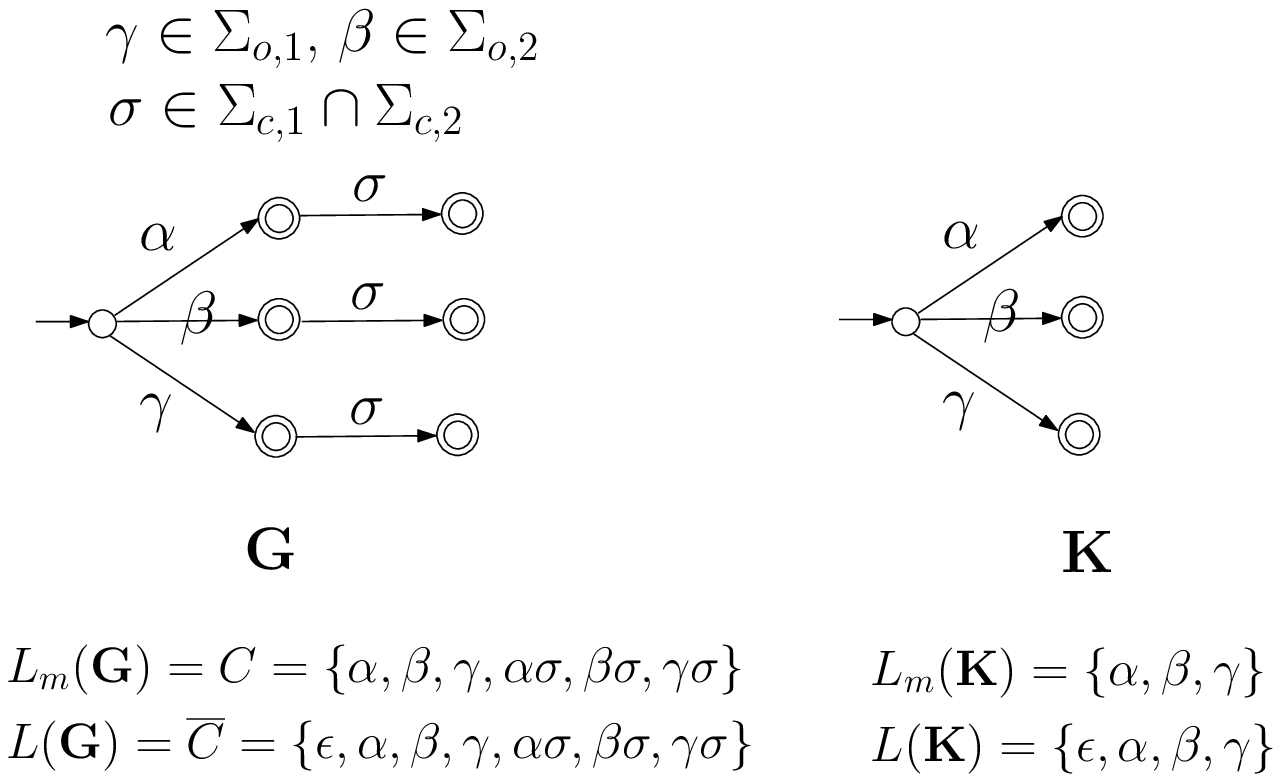}
  \caption{$L_m(\textbf{K})$ is relatively coobservable but not conormal. A straightforward calculation shows
  that $\left( P^{-1}_1 P_1 (\overline{K}) \cup P^{-1}_2 P_2 (\overline{K})
  \right) \cap L({\bf G}) = L({\bf G}) \supsetneqq \overline{K}$;
  hence $L_m(\textbf{K})$ is not conormal. On the other hand, by noting that the
  controllable event $\sigma$ is removed after strings $\alpha$, $\beta$, and $\gamma$, it is
  easily checked that $L_m(\textbf{K})$ is relatively observable
  with respect to both $P_1$ and $P_2$, and therefore is relatively
  coobservable.}
  \label{fig:relcoobs_norm}
\end{figure}

The reverse statement of Proposition~\ref{prop:relcoobs_norm} need
not be true; an example is displayed in
Fig.~\ref{fig:relcoobs_norm}.

\begin{rem}
A weak conormality concept was studied in \cite{TakaiUshio:02},
which is proved to be weaker than conormality and also preserved
under set union. However no algorithm is given to compute the
supremal element. Then in \cite{TakKumUsh:05}, weak conormality is
shown to be stronger than the ``strong local observability''. The
latter is the special case of our relative coobservability with the
largest possible ambient language $C = L_m({\bf G})$, hence the
strongest. Therefore we conclude that relative coobservability is
generally weaker than weak conormality.
\end{rem}


%
\section{Supremal Relatively Coobservable Sublanguage and Algorithms} \label{Sec3_SupRelCoobs}

First, we show that an arbitrary union of relatively coobservable
languages is again relatively coobservable. Let $\mathcal {I}$
denote the set of decentralized supervisors, and $P_i$ the natural
projection for each $i \in \mathcal {I}$.

\begin{prop} \label{prop:relcoobs_union}
Let $K_\alpha \subseteq C \subseteq L_m({\bf G})$, $\alpha \in
\mathcal {A}$ (some index set), be $\overline{C}$-coobservable. Then
$K = \bigcup\{K_\alpha \ |\ \alpha \in \mathcal {A}\}$ is also
$\overline{C}$-coobservable.
\end{prop}

\emph{Proof.} To prove that $K$ is $\overline{C}$-coobservable, we
show that $K$ is $\overline{C}$-observable with respect to $P_i$ for
each $i \in \mathcal {I}$. Let $i \in \mathcal {I}$, $s,s' \in
\Sigma^*$, $P_i(s) = P_i(s')$, $\sigma \in \Sigma_{c,i}$, $s \sigma
\in \overline{K}$, $s' \in \overline{C}$, and $s' \sigma \in
L(\textbf{G})$; it will be shown that $s' \sigma \in \overline{K}$.
Since $\overline{K} = \overline{\bigcup_{\alpha \in \mathcal {A}}
K_\alpha} = \bigcup_{\alpha \in \mathcal {A}} \overline{K_\alpha}$,
there exists $\alpha \in \mathcal {A}$ such that $s \sigma \in
\overline{K_\alpha}$. Since $K_\alpha$ is
$\overline{C}$-coobservable, it is $\overline{C}$-observable with
respect to $P_j$ for all $j \in \mathcal {I}$. In particular,
$K_\alpha$ is $\overline{C}$-observable with respect to $P_i$, and
thereby we derive that $s' \sigma \in \overline{K_\alpha}$. Finally
$s' \sigma \in \bigcup_{\alpha \in \mathcal {A}} \overline{K_\alpha}
= \overline{K}$. \hfill $\square$

In the proof to establish closure under union for relative
coobservability, it was essential that $K_\alpha$ ($\alpha \in
\mathcal {A}$) being $\overline{C}$-coobservable means that
$K_\alpha$ is $\overline{C}$-observable with respect to {\it all}
channels $P_j$, $j \in \mathcal {I}$. This confirms the importance
of using $\wedge$ in (\ref{eq:relcoobs1}) in the definition of
relative coobservability.

Now let $K \subseteq C \subseteq L_m({\bf G})$.  Whether or not $K$
is $\overline{C}$-coobservable, write
\begin{align} \label{eq:relcoobs-family}
\mathcal {O}(K, C) := \{ K' \subseteq K \ |\ K' \mbox{ is
$\overline{C}$-coobservable} \}
\end{align}
for the family of $\overline{C}$-coobservable sublanguages of $K$.
Note that the empty language $\emptyset$ is trivially
$\overline{C}$-coobservable, thus a member of $\mathcal {O}(K,C)$.
By Proposition~\ref{prop:relcoobs_union} we obtain that $\mathcal
{O}(K,C)$ has a unique supremal element sup$\mathcal {O}(K,C)$ given
by
\begin{align} \label{eq:sup-relcoobs}
\mbox{sup}\mathcal {O}(K,C) := \bigcup\{K' \ |\ K' \in \mathcal
{O}(K,C)\}.
\end{align}
This is the supremal $\overline{C}$-coobservable sublanguage of $K$.
We state these important facts about $\mathcal {O}(K,C)$ in the
following

\begin{thm} \label{thm:sup-relcoobs}
Let $K \subseteq C \subseteq L_m({\bf G})$. The set $\mathcal
{O}(K,C)$ is nonempty, and contains the supremal element
sup$\mathcal {O}(K,C)$ in (\ref{eq:sup-relcoobs}).
\end{thm}

%
%
%


Next we present an algorithm to compute sup$\mathcal {O}(K,C)$. The
idea is to apply the algorithm in \cite{CaiZhaWon:14TAC},
iteratively for each $P_i$ ($i \in \mathcal {I}$), to compute the
respective supremal relatively observable sublanguage. Let ${\bf
G}$, ${\bf C}$, and ${\bf K}$ be finite-state generators (as in
(\ref{eq:generator})) with marked languages $L_m({\bf G})$, $C$, and
$K$, respectively.

\emph{Algorithm 1:} Input $\textbf{G}$, ${\bf C}$, $\textbf{K}$, and
$P_i:\Sigma^* \rightarrow \Sigma^*_{o,i}$, $i \in \mathcal
{I}:=\{1,...,N\}$.

\noindent 1. Set ${\bf K}_0 := {\bf K}$.

\noindent 2. For $j \geq 0$, set ${\bf K}_{j,1} := {\bf K}_j$.

\noindent 3. For $i \geq 1$, apply the algorithm in
\cite{CaiZhaWon:14TAC} with inputs $\textbf{G}$, ${\bf K}_{j,i}$,
and $P_i$ to obtain ${\bf K}_{j,i+1}$ such that $L_m({\bf
K}_{j,i+1})$ is the supremal $\overline{C}$-observable sublanguage
of $L_m({\bf K}_{j,i})$ with respect to $P_i$. Proceed until ${\bf
K}_{j,N}$ is computed, and set it to be ${\bf K}_{j+1}$. If ${\bf
K}_{j+1} = {\bf K}_{j}$,\footnote{Here $=$ means that the two
generators are isomorphic \cite[Chapter~3]{SCDES}.} then output
$\textbf{K}^\uparrow := {\bf K}_{j+1}$. Otherwise, advance $j$ to
$j+1$ and go to Step~2.

Algorithm~1 terminates in finite steps, because the algorithm in
\cite{CaiZhaWon:14TAC} does so and removes states and/or transitions
from the finite-state generator ${\bf K}$. The complexity of
Algorithm~1 is exponential in the state size of $\textbf{K}$,
inasmuch as the algorithm in \cite{CaiZhaWon:14TAC} is of this
complexity.

\begin{thm} \label{thm:alg-relcoobs}
The output $\textbf{K}^\uparrow$ of Algorithm~1 satisfies $L_m({\bf
K}^\uparrow) = \mbox{sup} \mathcal {O}(K,C)$, the supremal
$\overline{C}$-coobservable sublanguage of $K$.
\end{thm}

{\it Proof.} First, it is guaranteed by Step~3 of Algorithm~1 that
$L_m({\bf K}^\uparrow)$ is $\overline{C}$-observable with respect to
$P_i$ for each $i \in \mathcal {I}$. Thus $L_m({\bf K}^\uparrow) \in
\mathcal {O}(K,C)$. It remains to prove that if $K' \in \mathcal
{O}(K,C)$, then $K' \subseteq L_m({\bf K}^\uparrow)$. To see this,
consider induction on the iterations $j=0,1,2,...$ (Step 2) of
Algorithm~1. Since $K' \subseteq K = L_m(\textbf{K})$, we have $K'
\subseteq L_m(\textbf{K}_0)$. Suppose now $K' \subseteq
L_m(\textbf{K}_j)$. Since $K'$ is $\overline{C}$-observable for all
$P_i$, no change will be made in the subsequent Step~3 by applying
the algorithm in \cite{CaiZhaWon:14TAC}. Therefore $K' \subseteq
L_m(\textbf{K}_{j+1})$, and eventually $K' \subseteq L_m({\bf
K}^\uparrow)$. \hfill $\square$

In practice we shall use Algorithm~1 as follows. Given a
(specification) language $K \subseteq L_m({\bf G})$, check if $K$ is
coobservable (polynomial algorithm available \cite{RudWon:95}). If
so, we stop. Otherwise apply Algorithm~1 to obtain the supremal
$\overline{K}$-coobservable sublanguage of $K$. Since relative
coobservability implies coobservability, the obtained supremal
sublanguage is also coobservable.


Now let us bring in control. Let $K \subseteq L_m({\bf G})$ be a
nonempty specification language. Since
$\overline{C}$-coobservability, controllability, and $L_m({\bf
G})$-closedness are all closed under set union, there exists the
supremal sublanguage of $K$ that satisfies these three properties.
Denote this supremal sublanguage by $K^\uparrow$; by
Proposition~\ref{prop:relcoobs_CPcoobs} (or
Proposition~\ref{prop:relcoobs_DAcoobs}), $K^\uparrow$ is
conjunctively (or disjunctively) coobservable, controllable, and
$L_m({\bf G})$-closed. Therefore, by \cite{RudWon:92} (resp.
\cite{YooLaf:02}) there exist decentralized supervisors
conjunctively (resp. disjunctively) synthesizing $K^\uparrow$.

We present an algorithm to compute $K^\uparrow$. Let ${\bf G}$ and
${\bf K}$ be finite-state generators (as in (\ref{eq:generator}))
with marked languages $L_m({\bf G})$ and $K$, respectively.

\emph{Algorithm 2:} Input $\textbf{G}$, $\textbf{K}$, and
$P_i:\Sigma^* \rightarrow \Sigma^*_{o,i}$, $i \in \mathcal {I}$.

\noindent 1. Set ${\bf K}_0 = {\bf K}$.

\noindent 2. For $j \geq 0$, apply the algorithm in \cite{WonRam:87}
with inputs $\textbf{G}$ and ${\bf K}_j$ to obtain ${\bf H}_j$ such
that $L_m({\bf H}_j)$ is the supremal controllable and $L_m({\bf
G})$-closed sublanguage of $L_m({\bf K}_j)$.

\noindent 3. Apply Algorithm~1 with inputs $\textbf{G}$, ${\bf
H}_j$, ${\bf H}_j$, and $P_i:\Sigma^* \rightarrow \Sigma^*_o$ ($i\in
\mathcal {I}$) to obtain ${\bf K}_{j+1}$ such that $L_m({\bf
K}_{j+1})$ is the supremal $L({\bf H}_j)$-coobservable sublanguage
of $L_m({\bf H}_j)$. If ${\bf K}_{j+1} = {\bf K}_{j}$, then output
$\textbf{K}^\uparrow = {\bf K}_{j+1}$. Otherwise, advance $j$ to
$j+1$ and go to Step~2.

Algorithm~2 terminates in finite steps, inasmuch as both algorithms
used in Steps~2 and 3 do so and both remove states and/or
transitions from the finite-state generator ${\bf K}$. The
complexity of Algorithm~2 is exponential in the state size of
$\textbf{K}$, because Algorithm~1 is of this complexity.


Note that in applying Algorithm~1 in Step~3 above, the ambient
language successively shrinks to the supremal controllable
sublanguage $L({\bf H}_j)$ computed at the immediately previous
Step~2. Using successively smaller ambient languages helps generate
less restrictive controlled behavior by discarding any strings
outside $L({\bf H}_j)$ that may be effectively prohibited by means
of control.


%
\section{Guideway} \label{Sec4_Guideway}



\begin{figure}[!t]
  \centering
  \includegraphics[width=0.45\textwidth]{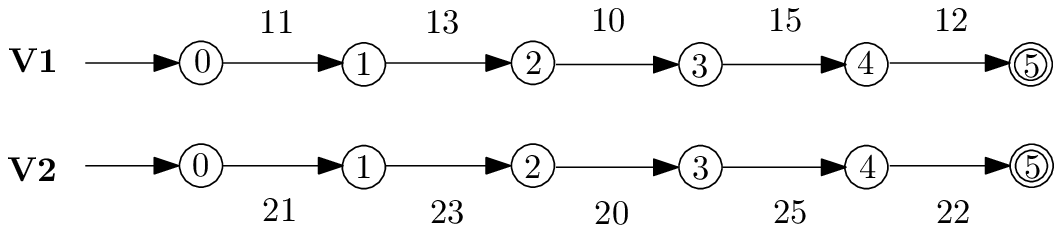}
  \caption{Vehicle generator models}
  \label{fig:vehicle}
\end{figure}

We demonstrate relative coobservability and Algorithm~2 with a
Guideway example, adapted from \cite[Section~6.6]{SCDES}. As
displayed in Fig.~\ref{fig:vehicle}, two vehicles, ${\bf V}_1$ and
${\bf V}_2$, use the Guideway simultaneously and travel from station
A (state 0) to B (state 5). The track between the two stations
consists of 4 sections (states 1, 2, 3, 4). The plant ${\bf G}$ to
be controlled is the synchronous product (e.g. \cite{SCDES}) ${\bf
G} = {\bf V}_1 || {\bf V}_2$, and the control specification is to
ensure that ${\bf V}_1$ and ${\bf V}_2$ never travel on the same
section of track simultaneously, i.e. ensure \emph{mutual exclusion}
of the state pairs $(j,j), j=1,...,4$. Let ${\bf K}$ be a generator
representing this specification.

\begin{figure}[!t]
  \centering
  \includegraphics[width=0.5\textwidth]{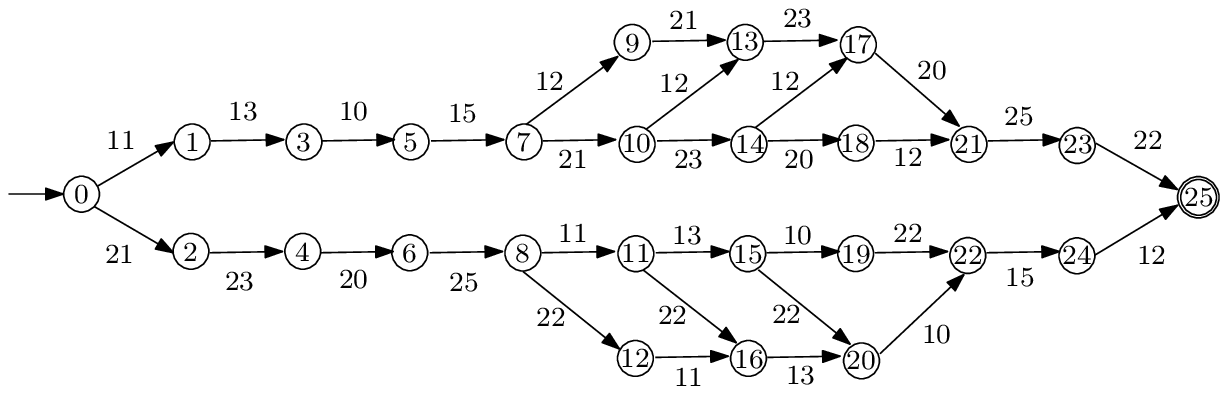}
  \caption{Supremal conormal, controllable, and $L_m({\bf G})$-closed
  sublanguage}
  \label{fig:SUPN}
\end{figure}

\begin{figure}[!t]
  \centering
  \includegraphics[width=0.5\textwidth]{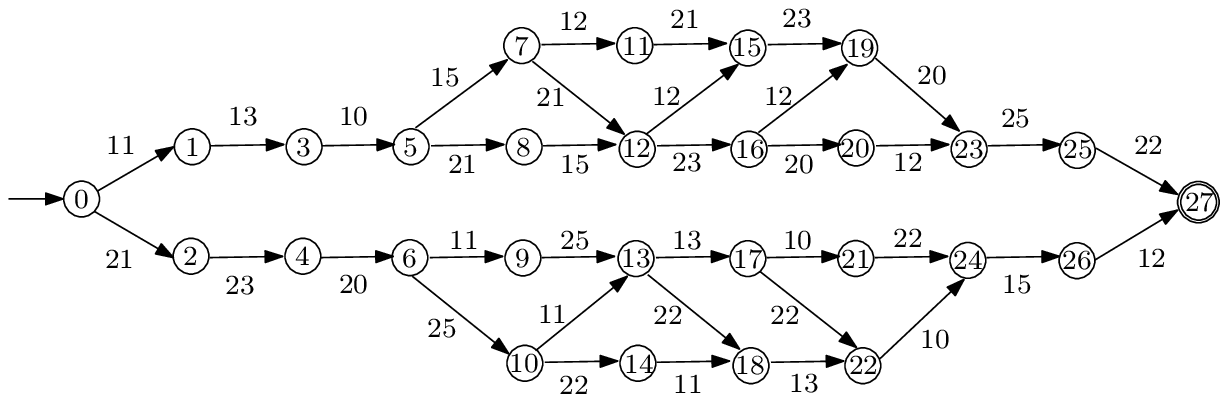}
  \caption{Supremal relatively coobservable, controllable, and $L_m({\bf G})$-closed sublanguage}
  \label{fig:supcoobs}
\end{figure}

\begin{figure}[!t]
  \centering
  \includegraphics[width=0.48\textwidth]{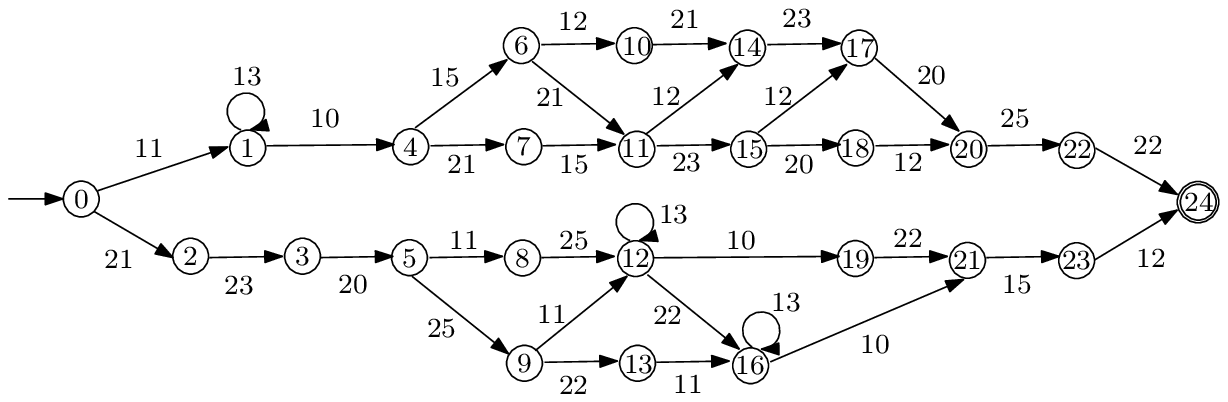}
  \caption{Decentralized supervisor ${\bf SUP}_1$. The unobservable controllable event $13$ is selflooped at
  those states where it is enabled.}
  \label{fig:UntimeDsup1}
\end{figure}

%

We consider the following decentralized supervisory control problem.
Suppose that there are two supervisors, with unobservable event
subsets $\Sigma_{uo,1}=\{13\}$, $\Sigma_{uo,2}=\{23\}$, and
controllable event subsets $\Sigma_{c,1}=\{11,13,23,15\}$,
$\Sigma_{c,2}=\{21,13,23,25\}$. The unobservable subsets
$\Sigma_{uo,i}$ define the corresponding natural projections $P_i$,
$i=1,2$, and the shared controllable events are 13, 23.

For comparison, we first compute the conormal,
controllable, and $L_m({\bf G})$-closed sublanguage, represented by
the generator in Fig.~\ref{fig:SUPN}. Then applying Algorithm~2, we
obtain the generator in Fig.~\ref{fig:supcoobs}, which represents
the supremal relatively coobservable, controllable, and $L_m({\bf
G})$-closed sublanguage. Observe that the relatively coobservable
controlled behavior is strictly more permissive than the conormal
counterpart. We next construct as in \cite{RudWon:92} the
corresponding two decentralized supervisors ${\bf SUP}_i$, with
$\Sigma_{uo,i}$ and $\Sigma_{c,i}$ ($i=1,2$); ${\bf SUP}_1$ is
displayed in Fig.~\ref{fig:UntimeDsup1} and ${\bf SUP}_2$ is
similar.

We explain a representative case of the control logic of ${\bf
SUP}_1$. If ${\bf SUP}_1$ observes that ${\bf V2}$ arrives at track
section~3 (i.e. after string 21.23.20), either it allows ${\bf V1}$
to enter section~1 (i.e. ${\bf SUP}_1$ enables its private event
11), or ${\bf V2}$ is allowed by ${\bf SUP}_2$ to move onto
section~4 (i.e. ${\bf SUP}_2$ enables its private event 25). When
the former occurs, ${\bf SUP}_1$ must prevent ${\bf V1}$ from
entering section~2 (i.e. ${\bf SUP}_1$ must disable the unobservable
event 13 at its state 8) because otherwise ${\bf V1}$ can thereafter
uncontrollably enter section 3 (event 10) and violate mutual
exclusion at section 3. Note that since event 13 is shared, in the
above case ${\bf SUP}_2$ must also disable 13. The above control
action is not possible for conormality, since disabling unobservable
events is not allowed. This is why relative coobservability achieves
strictly more permissive than conormality does.



%
\section{Conclusions} \label{Sec5_Concl}

We have studied the new concept of relative coobservability in
decentralized supervisory control of DES. We have proved that
relative coobservability is stronger than (any variations of)
coobservability, weaker than conormality, and closed under set
union. Moreover, we have presented an algorithm for computing the
supremal relatively coobservable (and controllable, $L_m({\bf
G})$-closed) sublanguage of a given language, and demonstrated the
result with a Guideway example.
In future work, we aim to apply relative coobservability in
decentralized control of large systems and follow the architectural
approach in \cite{FenWon:08}.


\bibliographystyle{IEEEtran}
\bibliography{DES,observable}

\begin{thebibliography}{10}
\providecommand{\url}[1]{#1}
\csname url@samestyle\endcsname
\providecommand{\newblock}{\relax}
\providecommand{\bibinfo}[2]{#2}
\providecommand{\BIBentrySTDinterwordspacing}{\spaceskip=0pt\relax}
\providecommand{\BIBentryALTinterwordstretchfactor}{4}
\providecommand{\BIBentryALTinterwordspacing}{\spaceskip=\fontdimen2\font plus
\BIBentryALTinterwordstretchfactor\fontdimen3\font minus
  \fontdimen4\font\relax}
\providecommand{\BIBforeignlanguage}[2]{{%
\expandafter\ifx\csname l@#1\endcsname\relax
\typeout{** WARNING: IEEEtran.bst: No hyphenation pattern has been}%
\typeout{** loaded for the language `#1'. Using the pattern for}%
\typeout{** the default language instead.}%
\else
\language=\csname l@#1\endcsname
\fi
#2}}
\providecommand{\BIBdecl}{\relax}
\BIBdecl

\bibitem{CaiZhaWon:14TAC}
K.~Cai, R.~Zhang, and W.~M. Wonham, ``Relative observability of discrete-event
  systems and its supremal sublanguages,'' \emph{IEEE Trans. Autom. Control},
  vol.~60, no.~3, pp. 659--670, 2015.

\bibitem{CaiZhaWon:13CDC}
------, ``On relative observability of discrete-event systems,'' in \emph{Proc.
  52nd IEEE Conf. Decision and Control}, Florence, Italy, 2013, pp. 7285--7290.

\bibitem{CaiZhaWon:14WODES}
------, ``On relative observability of timed discrete-event systems,'' in
  \emph{Proc. Workshop on Discrete-Event Systems}, Cachan, France, 2014, pp.
  208--213.

\bibitem{CaiZhaWon:ACC15}
------, ``On relative coobservability of discrete-event systems,'' in
  \emph{Proc. American Control Conference}, Chicago, IL, 2015, pp. 371--376.

\bibitem{SCDES}
W.~M. Wonham, ``Supervisory {C}ontrol of {D}iscrete-{E}vent {S}ystems,''
  {S}ystems Control Group, ECE Dept, University of Toronto, updated July 1,
  2015. Available online at http://www.control.toronto.edu/DES.

\bibitem{RudWon:92}
K.~Rudie and W.~M. Wonham, ``Think globally, act locally: decentralized
  supervisory control,'' \emph{IEEE Trans. Autom. Control}, vol.~37, no.~11,
  pp. 1692--1708, 1992.

\bibitem{Cieslak:88}
R.~Cieslak, C.~Desclaux, A.~S. Fawaz, and P.~Varaiya, ``Supervisory control of
  discrete-event processes with partial observations,'' \emph{IEEE Trans.
  Autom. Control}, vol.~33, no.~3, pp. 249--260, 1988.

\bibitem{YooLaf:02}
T.~S. Yoo and S.~Lafortune, ``A general architecture for decentralized
  supervisory control of discrete-event systems,'' \emph{Discrete Event Dynamic
  Systems}, vol.~12, no.~3, pp. 335--377, 2002.

\bibitem{YooLaf:04a}
------, ``Decentralized supervisory control with conditional decisions:
  supervisor existence,'' \emph{IEEE Trans. Autom. Control}, vol.~49, no.~11,
  pp. 1886--1904, 2004.

\bibitem{Tri:04letter}
S.~Tripakis, ``Undecidable problems of decentralized observation and control on
  regular languages,'' \emph{Information Processing Letters}, vol.~90, no.~1,
  pp. 21--28, 2004.

\bibitem{TakaiUshio:02}
S.~Takai and T.~Ushio, ``A modified normality condition for decentralized
  supervisory control of discrete event systems,'' \emph{Automatica}, vol.~38,
  no.~1, pp. 185--189, 2002.

\bibitem{TakKumUsh:05}
S.~Takai, R.~Kumar, and T.~Ushio, ``Characterization of co-observable languages
  and formulas for their super/sublanguages,'' \emph{IEEE Trans. Autom.
  Control}, vol.~50, no.~4, pp. 434--447, 2005.

\bibitem{KozWon:95}
P.~Kozak and W.~M. Wonham, ``Fully decentralized solutions of supervisory
  control problems,'' \emph{IEEE Trans. Autom. Control}, vol.~40, no.~12, pp.
  2094--2097, 1995.

\bibitem{RohLaf:03}
K.~Rohloff and S.~Lafortune, ``On the synthesis of safe control policies in
  decentralized control of discrete-event systems,'' \emph{IEEE Trans. Autom.
  Control}, vol.~48, no.~6, pp. 1064--1068, 2003.

\bibitem{LinWon:88Obs}
F.~Lin and W.~M. Wonham, ``On observability of discrete-event systems,''
  \emph{Inform. Sci.}, vol.~44, no.~3, pp. 173--198, 1988.

\bibitem{RudWon:95}
K.~Rudie and J.~C. Willems, ``The computational complexity of decentralized
  discrete-event control problems,'' \emph{IEEE Trans. Autom. Control},
  vol.~40, no.~7, pp. 1313--1319, 1995.

\bibitem{WonRam:87}
W.~M. Wonham and P.~J. Ramadge, ``On the supremal controllable sublanguage of a
  given language,'' \emph{SIAM J. of Control and Optimization}, vol.~25, no.~3,
  pp. 637--659, 1987.

\bibitem{FenWon:08}
L.~Feng and W.~M. Wonham, ``Supervisory control architecture for discrete-event
  systems,'' \emph{IEEE Trans. Autom. Control}, vol.~53, no.~6, pp. 1449--1461,
  2008.

\end{thebibliography}


\end{document}